\documentclass[10pt,draft]{dis03}
\usepackage{epsfig,amsmath}

\newcommand{\be}{\begin{eqnarray}}
\newcommand{\ee}{\end{eqnarray}}

\begin{document}

\title{BASICS OF THE COLOR GLASS CONDENSATE
\thanks{Contribution to the 11th International 
Workshop on Deep--Inelastic Scattering (DIS03), St. Petersburg, April 2003}
}

\author{Edmond Iancu\\
Service de Physique 
Th\'eorique, CEA Saclay, France\\
91191 Gif--sur--Yvette cedex, France\\
E-mail: eiancu@cea.fr}

\maketitle

\begin{abstract}
\noindent I review basic concepts of the effective theory for the
color glass condensate which describes the high--energy limit of  QCD interactions.
\end{abstract}

The physics of the color glass condensate [1--3] 
covers and unifies under the same
banner topics which have originally appeared and developed under various
names --- ``small--$x$ physics'', ``BFKL evolution'',
``unitarity corrections'', ``parton saturation'',
``multiple pomeron exchanges'', ``higher twist effects'', etc. ---, but which
are modernly understood as manifestations or consequences of the same basic
physical mechanism: a change in the form of gluonic matter in the hadron
wavefunction at small--$x$. 
This change can be visualized as a ``critical line'' which divides the kinematical
plane for deep inelastic scattering into
two regions (see Fig. \ref{phase-diagram}): a low density region at 
high $Q^2$ (for a given value of $x$, or $\tau\equiv \ln(1/x)$), in which
parton densities evolve according to linear evolution equations (DGLAP or BFKL)
and grow rapidly with $1/x$, and a high density regime at relatively
low\footnote{With $Q^2\gg \Lambda_{QCD}^2$, though: in what follows, I shall always
assume weak coupling.} $Q^2$, where the parton densities saturate because of the
large non--linear effects, and the gluons form a {\it condensate}.
This is a high--density state characterized by an intrinsic scale, 
the saturation momentum $Q_s(x)$, 
and by large occupation numbers, of order $1/\alpha_s$,
for the gluonic modes with momentum less than or equal to $Q_s$.
The saturation momentum is the typical momentum of the saturated gluons,
and grows rapidly with the energy, as a power of $1/x\,$.
The saturation line $Q^2=Q_s^2(x)$ is the separating line in Fig. \ref{phase-diagram}.
Note that the transition across this line is rather smooth, and should
not be thought of as a phase transition in the sense of thermodynamics: to my
knowledge, no quantity becomes discontinuous at this line. The smooth character
of the transition is best demonstrated by the fact that a qualitative change of
behaviour appears already {\it before} crossing this line, namely, in the
region $Q^2>Q^2_s(x)$, where one sees some more structure
in  Fig. ~\ref{phase-diagram}: 
\begin{figure}[!thb]
\vspace*{7.cm}
\begin{center}
\includegraphics{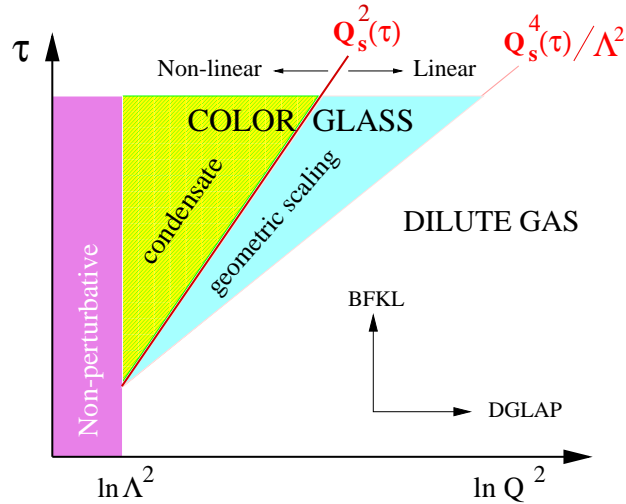}
%\vspace*{1cm}
\caption[*]{Domains for evolution
in the $\tau-\ln Q^2$ plane, with $\tau=\ln(1/x)$.}
\label{phase-diagram} 
\end{center}
\vspace*{-.8cm}
\end{figure}

In addition to the truly dilute regime at very high $Q^2$, where the leading twist
approximation is excellent and DGLAP equation applies, there is also an intermediate
regime in between the saturation line and the ``geometric scaling'' line 
\cite{geometric,SCALING,KI03}
where the gluon density is large enough to entail strong correlations. These correlations
depend upon $Q^2$ and $x$ according to {\it scaling laws} characteristic for a liquid:
these are powers of $Q^2/Q^2_s(x)$, with the exponents determined
by the BFKL equation. In particular, this behaviour leads to DIS structure functions
which violate {\it strongly} Bjorken scaling (i.e., which show a power--law dependence
upon  $Q^2$) \cite{KI03}. 
This intermediate region is where BFKL evolution truly applies. Note that,
in order to reach higher and higher energies within the BFKL description, one has to
simultaneously move up to larger and larger values of $Q^2$, in such a way to remain
in the linear regime. Thus, strictly speaking, BFKL equation is not the right tool
to study the high--energy limit of QCD, defined as the limit
$x\to 0$ at fixed $Q^2$. Nevertheless, BFKL provides the right approach towards 
saturation, and thus determines the scale which plays the decisive role for scattering
at very high energies: the saturation momentum. % $Q_s(x)$.

At all the points on the left of the ``geometric scaling'' line in
in  Fig. \ref{phase-diagram} (this includes the BFKL regime and the
gluon condensate at saturation), the gluons can be described as a {\it color glass}.
This is a glass because
the small--$x$ gluons are typically radiated by color sources (partons)
with much larger values of $x$, and thus larger rapidities, whose internal dynamics
is slowed down by Lorentz time dilation. Accordingly, the radiated gluons evolve
very slowly relative to natural time scales, so an external probe --- like
the virtual photon in DIS at small--$x$ --- ``sees'' only a ``frozen'' configuration
of these gluons, which can be any of the configurations allowed by the dynamics
of their fast moving sources. Physical observables like cross--sections
are then obtained by averaging over all such configurations, with some
``weight function'' (a functional of the color charge density) which 
describes the spatial distribution of the sources.

Besides being physically intuitive, this picture is interesting in that it offers
a natural framework for the inclusion of non--linear effects 
like gluon recombination:
These effects appear either as non--linearities in the classical Yang--Mills
equations which relate the color fields to their sources, or as correlations
among these sources, encoded in the weight function.
These correlations are built up in the course of
the quantum evolution towards small $x$. This involves a renormalization group
analysis \cite{JKLW97,PI}, 
in which quantum fluctuations are `integrated out' in layers of rapidity
and in the presence of the color fields radiated by the sources constructed in 
previous steps. This is done in the leading logarithmic approximation with respect 
to $\ln(1/x)$, to preserve the hierarchy of time scales.
The result of this calculation \cite{PI} is a functional differential
equation which governs the evolution of the
weight function with increasing $\tau=\ln 1/x$. 

The fact that this is
a functional equation means that it is equivalent to an infinite hierarchy of ordinary
evolution equations for $n$--point correlation functions. In 
particular, if these equations are written for Wilson lines operators
(the relevant operators for scattering at high energy), they turn
out to be the same as the equations originally derived by Balitsky 
\cite{B}, by using the operator--product expansion near the light--cone. 
In fact, within the space of Wilson lines, the functional equation 
for the color glass is equivalent  \cite{BIW} to a similar equation deduced
by Weigert \cite{W} from an analysis of Balitsky's equations.

In the low density regime at $Q > Q_s(x)$, where the color fields are weak,
the non--linear effects can be neglected. Then, the equation for the 2--point
function reduces to the BFKL equation for the gluon distribution \cite{JKLW97,PI}. 
By itself, this equation does not describe saturation, but
can be used to determine the saturation scale from the condition
that the gluon density becomes of order $1/\alpha_s$ when
$Q \sim Q_s(x)$. This yields :
%\be\label{Qs}
$Q_s^2(\tau)\,=\,Q^2_0\,\,{\rm e}^{\,\lambda\tau}$,
%\,,\ee
where only the exponent $\lambda$ is under control.
A leading--order BFKL calculation with fixed coupling constant
gives $\lambda = c (\alpha_s N_c/\pi)$ with $N_c=3$ and $c=4.88...$
\cite{GLR,AM99,SCALING}.
This is of order one for $\alpha_s \sim 0.2$, which is far too large to
agree with the phenomenology at HERA within the ``saturation model'' \cite{GBW99,geometric}.
Remarkably, however, a recent calculation \cite{DT02} using the
NLO BFKL formalism reduces this value to
$\lambda \approx 0.3$, in good agreement with the fits in Refs.
\cite{GBW99,geometric}.

More generally, it has been recently  shown \cite{IM03} that, in
the weak field and large--$N_c$ limits, the functional evolution equation 
for the color glass \cite{PI} is equivalent
to Mueller's color dipole picture of the onium wavefunction \cite{AM94}.

But the most interesting regime, of course, is the high--density regime at $Q
\le Q_s(\tau)$. In this regime, the color fields are strong --- the typical
field strength is of order $1/g$ ---, so the non--linear effects must be included
exactly. Then, the functional evolution equation is too complicated to be
solved exactly (except maybe via numerical simulations), but mean field 
approximations have been constructed \cite{SAT}, with interesting physical
conclusions: First, the gluon phase--space density is explicitly seen to saturate 
at a value of order  $1/\alpha_s$, as expected. Second, the saturated gluons
are spatially correlated with each other in such a way to shield their
color charges over a transverse area $\sim 1/Q^2_s$. This results in
a smoother spectrum at low transverse momenta, which eliminates the infrared
problems of perturbation theory. For most purposes,  the saturation momentum 
acts effectively as an infrared cutoff. This justifies a posteriori the use
of perturbation theory. Third, when applied to deep inelastic scattering
(and also to high--energy onium--onium scattering \cite{AM94,IM03}),
saturation ensures the unitarization of the $S$--matrix at fixed impact parameter
\cite{KI03}.

These conclusions are corroborated by various, analytic and numerical,
investigations of the Kovchegov equation \cite{K}, which is a simple 
(since closed) non--linear evolution equation for the  scattering 
amplitude, and may be seen as a special approximation to the first equation in
the hierarchy by Balitsky. This equation is a convenient laboratory to study
conceptual and phenomenological consequences of saturation, with many recent
applications. I refer to \cite{IV03} for a recent review and more references,
and also to several contributions to this meeting which have addressed,
more or less directly, this topic \cite{DIS03}.

To summarize, the color glass effective theory provides an unified description
of the BFKL regime, of the physics of saturation, and of the transition between
these two regimes. It also applies \cite{MV94}
to large nuclei ($A\gg 1$), which involve high gluon densities even at
not so small values of $x$ (say, $x\sim 10^{-2}$, as in the experiences at RHIC),
because of the many tree--level color sources: 
the $3 A$ valence quarks. 
There have been many interesting applications of this formalism
to the phenomenology of heavy ion collisions at RHIC, 
that I have no time to discuss here, but for which I
refer to the recent review paper \cite{IV03}.


\vspace*{-.3cm}


\begin{thebibliography}{0}

\bibitem{IV03}
For a recent review, see E.~Iancu and R. Venugopalan,
{\it The Color Glass Condensate and High Energy Scattering in QCD}, 
hep-ph/0303204. 
% Published  in {\it Quark-Gluon Plasma 3}, 
%Eds. R. C. Hwa and X.-N. Wang, World Scientific, 2003. 


\bibitem{MV94}
L. McLerran and R.~Venugopalan, {\it Phys.\ Rev.}\ {\bf D49} (1994) 2233;
{\it ibid.} {\bf 49} (1994) 3352; {\it ibid.} {\bf 50} (1994) 2225.


\bibitem{PI}
E.~Iancu, A.~Leonidov, and L.~McLerran,
 {\it Nucl. Phys.}~{\bf A692} (2001) 583;
{\it Phys. Lett.} {\bf B510} (2001) 133;
E.~Ferreiro, E.~Iancu, A.~Leonidov and L.~McLerran, 
{\it Nucl. Phys.} {\bf A703} (2002) 489.


\bibitem{geometric}
A. Sta\'sto, K.~Golec-Biernat, J.~Kwieci\'nski, 
{\it Phys. Rev. Lett.} {\bf 86} (2001) 596.


\bibitem{SCALING}
E.~Iancu, K. Itakura, and L. McLerran, {\it Nucl. Phys.} {\bf A708} (2002) 327;\\
A. H. Mueller and D.N. Triantafyllopoulos, 
{\it Nucl. Phys.} {\bf B640} (2002) 331.

\bibitem{KI03}
For more details see K.~Itakura, contribution to DIS03, these proceedings.


\bibitem{JKLW97}
J.~Jalilian-Marian, A.~Kovner, A.~Leonidov and  H.~Weigert,
{\it Nucl.\ Phys.}\ {\bf B504} (1997) 415;
{\it Phys.\ Rev.}\ {\bf D59} (1999) 014014.

\bibitem{B}
I.~Balitsky, {\it Nucl.\ Phys.}\ {\bf B463} (1996) 99;
%{\it High-energy QCD and Wilson lines}, 
hep-ph/0101042.

\bibitem{BIW}J.~P.~Blaizot, E.~Iancu and H.~Weigert,
{\it Nucl.\ Phys.}  {\bf A713} (2003) 441.

\bibitem{W}
H. Weigert, {\it Nucl. Phys.} {\bf A703} (2002) 823.

\bibitem{GLR}   L.V.~Gribov, E.M.~Levin, and M.G.~Ryskin, {\it Phys.
Rept. } {\bf 100} (1983) 1.
 


\bibitem{AM99} A. H. Mueller, {\it Nucl. Phys.} {\bf B558} (1999) 285.



\bibitem{DT02}
D.N. Triantafyllopoulos, {\it Nucl. Phys.} {\bf B648} (2002) 293.

\bibitem{GBW99}
K. Golec-Biernat and M. W\"usthoff, {\it Phys. Rev.} {\bf D59} (1999)
014017. %{\bf D60} (1999) 114023.
%{\it Eur. Phys. J.} {\bf C20} (2001) 313.

\bibitem{IM03}
E.~Iancu and A. H. Mueller, hep-ph/0308315.

\bibitem{AM94} A. H. Mueller, {\it Nucl. Phys.} {\bf B415} (1994) 373;
{\it ibid.} {\bf B437} (1995) 107.

\bibitem{SAT}
E.~Iancu and L.~McLerran, {\it Phys. Lett.} {\bf B510} (2001) 145;\\
E.~Iancu, K.~Itakura and L.~McLerran,
{\it Nucl. Phys.} {\bf A724} (2003) 181.

\bibitem{K}  Yu. V. Kovchegov, {\it Phys. Rev.} 
{\bf D60} (1999) 034008; {\bf D61} (2000) 074018.


\bibitem{DIS03}
J. Bartels,
H. Kowalski, M. Kozlov, E. Levin, M. Lublinsky, A.H. Mueller, S. Munier, M.G.~Ryskin,
A. Sta\'sto, these proceedings.

\end{thebibliography}
\end{document}